\documentstyle[12pt]{article}
\huge
 
\def\setzero{\setcounter{equation}{0}}

\newcounter{eqalph}

\begin{document}

\baselineskip 18pt

\def \sech{{\rm sech}}
\def \tanh{{\rm tanh}}
\def \cn{{\rm cn}}
\def \sn{{\rm sn}}
\def\bm#1{\mbox{\boldmath $#1$}}
\newfont{\bg}{cmr10 scaled\magstep4}
\newcommand{\bzr}{\smash{\hbox{\bg 0}}}
\newcommand{\bzl}{%
   \smash{\lower1.7ex\hbox{\bg 0}}}
\title{Unitary Matrix Models  and Painlev\'{e} III} 
\date{\today}
\author{ Masato {\sc Hisakado}  
\\
\bigskip
\\
{\small\it Graduate School of Mathematical Sciences,}\\
{\small\it University of Tokyo,}\\
{\small\it 3-8-1 Komaba, Megro-ku, Tokyo, 113, Japan}}
\maketitle

\vspace{20 mm}

Abstract

We discussed  the  full unitary matrix models 
from the   view points of integrable equations and 
string equations.
Coupling  the Toda equations  and the string equations, 
we derive a special case of the  Painlev\'{e} III equation.
 From  the Virasoro constrains,
 we can use the radial coordinate.
The relation between $t_{1}$ and $t_{-1}$  is like the complex conjugate.

\newpage

\section{Introduction}

Models  of the symmetric unitary matrix model are 
solved exactly in the double scaling limit,
 using orthogonal polynomials 
on a circle.\cite{p}
The partition function is the form
$\int dU\exp\{-\frac{N}{\lambda}{\rm tr} V(U)\}$, 
where $U$ is an $N\times N$ unitary matrix 
and tr$V(U)$ is some well defined function of $U$.
When $V(U)$ is the self adjoint  we call the model symmetric.\cite{b}
The simplest case is given by $V(U)=U+U^{\dag}$.
This unitary models has been studied in 
connection with the large-$N$ approximation 
to QCD in two dimensions.({\it one-plaquette model})\cite{gw}
For this model ``string equation'' is the   
Painlev\'{e} II equation.
``Discrete string equation'' is called the discrete 
Painlev\'{e} II equation.\cite{g}
When $V(U)$ is the anti-self adjoint,  we call the model anti-symmetric
 model.
The simplest case is given by $V(U)=U-U^{\dag}$. 
This is the theta term in the two-dimensional QCD.\cite{kts}
It has the topological meaning.
The full non-reduced unitary model was first discussed 
in \cite{m2}.
The full unitary model can be embedded in the 
two-dimensional Toda Lattice hierarchy.

In this letter we shall try to reformulate 
the full unitary matrix model 
from the view points of 
integrable equations and string equations.
These  two view points are closely 
connected each other 
to describe this model.
We unify these view points 
and clarify  a relation 
between these view
 points.

This letter is organized as follows.
In the section 2
we present   discrete string equations 
for  the full unitary matrix model.
Here we consider only the simplest case.
From  the Virasoro constraints,
 a relation between  times  $t_{1}$ and $t_{-1}$
is like complex conjugate.
Because of  this symmetry, we can use the radial coordinate.
In the section 3
 coupling  the Toda equation and the discrete string equations, 
we obtain  the special Painlev\'e III equation.
In the section 4
 we consider the reduced models, the symmetric  and 
the anti-symmetric model.
From the symmetric  and the anti-symmetric model  we can obtain the modified 
Volterra equation and the discrete nonlinear Schr\"{o}dinger 
equation respectively.
We study the relation  of the symmetric and the anti-symmetric model.
In a special case, we can transform the symmetric model
 into the anti-symmetric model.
Using this map, we can obtain B\"{a}cklund transformation 
of the  modified 
Volterra equation and the discrete nonlinear Schr\"{o}dinger 
equation.
The last section is devoted to concluding remarks.

\setzero

\section{Unitary Matrix model}

It is well known that  the partition function $\tau_{n}$ 
of the unitary matrix model  can be presented  as 
a product  of norms  of the biorthogonal polynomial system.
Namely, let us introduce  a scalar product of the form 
\begin{equation}
<A,B>=\oint\frac{d\mu(z)}{2\pi i z}
A(z)B(z^{-1}).
\end{equation}
where 
\begin{equation}
d\mu(z)=dz \exp\{-\sum_{m>0}(t_{m}z^{m}+t_{-m}z^{-m})\}
\end{equation}
Let us define  the system of the polynomials  biorthogaonal 
with respect to this scalar product 
\begin{equation}
<\Phi_{n},\Phi_{k}^{*}>=h_{k}\delta_{nk}.
\label{or}
\end{equation}
Then, the partition function  $\tau_{n}$ of the unitary matrix model 
is equal to the product of $h_{n}$'s:
\begin{equation}
\tau_{n}=\prod_{k=0}^{n-1}h_{k},\;\;\;\tau_{0}=1.
\end{equation}
The polynomials are normalized as follows
(we should stress that superscript `*' does not mean the 
complex conjugation):
\begin{equation}
\Phi_{n}=z^{n}+\cdots+S_{n-1},\;\;\Phi_{n}^{*}
=z^{n}+\cdots+S_{n-1}^{*},\;\;
S_{-1}=S_{-1}^{*}\equiv 1.
\label{2.4}
\end{equation}
Now it is easy to show that these polynomials satisfy the following
recurrent relations, 
\begin{eqnarray}
\Phi_{n+1}(z)&=&z\Phi_{n}(z)+S_{n}z^{n}\Phi_{n}^{*}(z^{-1}),
\nonumber \\
\Phi_{n+1}^{*}(z^{-1})&=&z^{-1}\Phi_{n}^{*}(z^{-1})
+S_{n}^{*}z^{-n}\Phi_{n}(z),
\end{eqnarray}
and
\begin{equation}
\frac{h_{n+1}}{h_{n}}=1-S_{n}S_{n}^{*}.
\end{equation}
Note that $h_{n}$, $S_{n}$, $S_{n}^{*}$ ,$\Phi_{n}(z)$
and  $\Phi_{n}^{*}$ depend parametrically  on
$t_{1},t_{2},\cdots,$ and $t_{-1},t_{-2},\cdots,$  but for convenience 
of notation we suppress this dependence.
Hereafter we call $t_{1},t_{2},\cdots,$ and $t_{-1},t_{-2},\cdots,$
time variables.

Using (\ref{or}) and integration by parts, we can obtain next relations:
\begin{eqnarray}
&-&\oint\frac{d\mu(z)}{2\pi i z} V'(z)\Phi_{n+1}(z)\Phi_{n}^{*}(z^{-1})
\nonumber\\
=& & -\oint\frac{d\mu(z)}{2\pi i z}\frac{\partial \Phi_{n+1}(z)}
{\partial z} \Phi_{n}^{*}(z^{-1})
-\oint\frac{d\mu(z)}{2\pi i z} \Phi_{n+1}(z)
\frac{\partial \Phi_{n}^{*}(z^{-1})}{\partial z}
+\oint\frac{d\mu(z)}{2\pi i z}\frac{\Phi_{n+1}(z)\Phi_{n}^{*}}{z}
\nonumber \\
=& & (n+1)(h_{n+1}-h_{n}),\label{se1}
\end{eqnarray}
and 
\begin{eqnarray}
& &\oint\frac{d\mu(z)}{2\pi i z} z^{2}V'(z)\Phi_{n+1}^{*}(z^{-1})\Phi_{n}(z)
\nonumber\\
&=& \oint\frac{d\mu(z)}{2\pi i z} z^{2}\frac{\partial \Phi_{n+1}^{*}
(z^{-1})}{\partial z}\Phi_{n}(z)
+\oint\frac{d\mu(z)}{2\pi i z} z^{2}\Phi_{n+1}^{*}(z^{-1})
\frac{\partial \Phi_{n}(z)}{\partial z}
+\oint\frac{d\mu(z)}{2\pi i z} z\Phi_{n+1}^{*}(z^{-1})\Phi_{n}(z)
\nonumber \\
&=& (n+1)(h_{n+1}-h_{n}).
\label{se2}
\end{eqnarray}
(\ref{se1}) and (\ref{se2}) are string equations for 
the full unitary matrix model.

If $t_{1}$ and $t_{-1}$
are free variables  while 
$t_{2}=t_{3}=\cdots=0$ and $t_{-2}=t_{-3}=\cdots=0$,
 (\ref{se1}) and (\ref{se2})
 become
\begin{equation}
 (n+1)S_{n}S_{n}^{*}=t_{-1}(S_{n}S_{n+1}^{*}+S_{n}^{*}S_{n-1})
(1-S_{n}S_{n}^{*}),
\label{edp1}
\end{equation}
\begin{equation}
 (n+1)S_{n}S_{n}^{*}=t_{1}(S_{n}^{*}S_{n+1}+S_{n}S_{n-1}^{*})
(1-S_{n}S_{n}^{*}).
\label{edp2}
\end{equation}

Next we introduce a useful relation.
Using (\ref{or}) and integration by parts,  we can show
\begin{eqnarray}
& &\oint\frac{d\mu(z)}{2\pi i z}zV'(z)\Phi_{n}(z)\Phi_(n)^{*}(z^{-1})
\nonumber \\
&=&
\oint\frac{d\mu(z)}{2\pi i z}z\frac{\partial \Phi_{n}(z)}
{\partial z}\Phi_{n}^{*}(z^{-1})
+\oint\frac{d\mu(z)}{2\pi i z}z\Phi_{n}(z)
\frac{\partial \Phi_{n}^{*}(z^{-1})}{\partial z}
\nonumber \\
&=& nh_{n}-nh_{n}=0.
\label{1}
\end{eqnarray}
This corresponds to  the Virasoro constraint:\cite{b}
\begin{equation}
L_{0}^{-cl}
=\sum_{k=-\infty}^{\infty}
kt_{k}\frac{\partial}{\partial t_{n}}.
\end{equation}
This relation constrains a symmetry like   complex conjugate
between $t_{k}$ and $t_{-k}$.
It is important in the next section.
If we set that $t_{1}$ and $t_{-1}$
are free variables  while 
$t_{2}=t_{3}=\cdots=0$ and $t_{-2}=t_{-3}=\cdots=0$,
 from (\ref{1}) we get
\begin{equation}
t_{1}S_{n}S_{n-1}^{*}=t_{-1}S_{n}^{*}S_{n-1}.
\label{2}
\end{equation}
Using (\ref{2}), (\ref{edp1}) and (\ref{edp2}) can be written
\begin{equation}
 (n+1)S_{n}=(t_{1}S_{n+1}+t_{-1}S_{n-1})
(1-S_{n}S_{n}^{*}),
\label{edp3}
\end{equation}
\begin{equation}
 (n+1)S_{n}^{*}=(t_{-1}S_{n+1}^{*}+t_{1}S_{n-1}^{*})
(1-S_{n}S_{n}^{*}).
\label{edp4}
\end{equation}

\setzero
\section{Toda equation and String equations}

Using the orthogonal conditions, it is also possible to obtain the
equations which describe the time dependence of $\Phi_{n}(z)$ 
and $\Phi_{n}^{*}(z)$. 
Namely, differentiating (\ref{or}) with  respect to  times
 $t_{1}$ and $t_{-1}$ gives the following evolution equations:
\begin{equation}
\frac{\partial \Phi_{n}(z)}{\partial t_{1}}
=-\frac{S_{n}}{S_{n-1}}\frac{h_{n}}{h_{n-1}}
(\Phi_{n}(z)-z\Phi_{n-1}),
\end{equation}
\begin{equation}
\frac{\partial \Phi_{n}(z)}{\partial t_{-1}}
=\frac{h_{n}}{h_{n-1}}\Phi_{n-1}(z),
\end{equation}
\begin{equation}
\frac{\partial \Phi_{n}^{*}(z^{-1})}{\partial t_{1}}
=\frac{h_{n}}{h_{n-1}}\Phi_{n-1}^{*}(z^{-1}),
\end{equation}
\begin{equation}
\frac{\partial \Phi_{n}^{*}(z^{-1})}{\partial t_{-1}}
=-\frac{S_{n}^{*}}{S_{n-1}^{*}}\frac{h_{n}}{h_{n-1}}
(\Phi_{n}^{*}(z^{-1})-z^{-1}\Phi_{n-1}^{*}),
\end{equation}
The compatibility condition gives the following nonlinear evolution equations:
\begin{equation}
\frac{\partial S_{n}}{\partial t_{1}}=-S_{n+1}\frac{h_{n+1}}{h_{n}},
\;\;\;
\frac{\partial S_{n}}{\partial t_{-1}}=S_{n-1}\frac{h_{n+1}}{h_{n}},
\label{11}
\end{equation}
\begin{equation}
\frac{\partial S^{*}_{n}}{\partial t_{1}}=S_{n+1}^{*}\frac{h_{n+1}}{h_{n}},
\;\;\;
\frac{\partial S^{*}_{n}}{\partial t_{-1}}=-S_{n-1}^{*}\frac{h_{n+1}}{h_{n}},
\label{22}
\end{equation}
\begin{equation}
\frac{\partial h_{n}}{\partial t_{1}}=S_{n}S_{n-1}^{*}h_{n},
\;\;\;
\frac{\partial h_{n}}{\partial t_{-1}}=S_{n}^{*}S_{n-1}h_{n},
\end{equation}
Here we define  $a_{n}$, $b_{n}$ and $b_{n}^{*}$:
\begin{equation}
a_{n}\equiv 1-S_{n}S_{n}^{*}=\frac{h_{n+1}}{h_{n}},
\end{equation}
\begin{equation}
b_{n}\equiv S_{n}S_{n-1}^{*},
\end{equation}
\begin{equation}
b_{n}^{*}\equiv S_{n}^{*}S_{n-1}.
\end{equation}
Notice that from  the definitions  $a_{n}$, $b_{n}$ and $b_{n}^{*}$
satisfy the following identity:
\begin{equation}
b_{n}b_{n}^{*}=(1-a_{n})(1-a_{n-1}).
\label{*1}
\end{equation}
 It can be shown  using (\ref{2}) that 
\begin{equation}
t_{1}b_{n}=t_{-1}b_{n}^{*}.
\label{*2}
\end{equation}

In terms of  $a_{n}$, $b_{n}$ and $b_{n}^{*}$,
 (\ref{11}) and (\ref{22}) become the two-dimensional  Toda  equations:
\begin{equation}
\frac{\partial a_{n}}{\partial t_{1}}
=a_{n}(b_{n+1}-b_{n}),\;\;\;
\frac{\partial b_{n}}{\partial t_{-1}}
=a_{n}-a_{n-1},
\label{t1}
\end{equation}
and 
\begin{equation}
\frac{\partial a_{n}}{\partial t_{-1}}
=a_{n}(b_{n+1}^{*}-b_{n}^{*}),\;\;\;
\frac{\partial b_{n}^{*}}{\partial t_{1}}
=a_{n}-a_{n-1}.
\label{t2}
\end{equation} 

Using $a_{n}$, $b_{n}$ and $b_{n}^{*}$,
 we  rewrite (\ref{edp1}) and (\ref{edp2})
\begin{equation}
\frac{n+1}{t_{-1}}\frac{1-a_{n}}{a_{n}}
=b_{n+1}^{*}+b_{n}^{*}.
\label{s2}
\end{equation}
and 
\begin{equation}
\frac{n+1}{t_{1}}\frac{1-a_{n}}{a_{n}}
=b_{n+1}+b_{n},
\label{s1}
\end{equation}

From (\ref{t1}) and  (\ref{s1}) 
we eliminate $b_{n+1}$,
\begin{equation}
2b_{n}=\frac{1}{a_{n}}[\frac{n+1}{t_{1}}(1-a_{n})-
\frac{\partial a_{n}}{\partial t_{1}}].
\label{ww1}
\end{equation}
In the same way, from (\ref{t2}) and  (\ref{s2}) 
we eliminate $b_{n+1}^{*}$,
\begin{equation}
2b_{n}^{*}=\frac{1}{a_{n}}[\frac{n+1}{t_{-1}}(1-a_{n})-
\frac{\partial a_{n}}{\partial t_{-1}}].
\label{ww2}
\end{equation}
Using (\ref{*1}) and (\ref{*2}),
(\ref{t1}) and (\ref{t2}) can be written 
\begin{equation}
\frac{\partial b_{n}}{\partial t_{-1}}
=
(a_{n}-1)+\frac{t_{1}}{t_{-1}}\frac{b_{n}^{2}}{1-a_{n}},
\label{w1}
\end{equation}
\begin{equation}
\frac{\partial b_{n}^{*}}{\partial t_{1}}
=
(a_{n}-1)+\frac{t_{-1}}{t_{1}}\frac{(b_{n}^{*})^{2}}{1-a_{n}}.
\label{w2}
\end{equation}
Using  (\ref{ww1}) and (\ref{w1})
to eliminate $b_{n}$
we obtain a second order ODE for $a_{n}$
\begin{eqnarray}
\frac{\partial ^{2}a_{n}}{\partial t_{1}\partial t_{-1}}
&=&
\frac{n+1}{t_{-1}a_{n}}\frac{\partial a_{n}}{\partial t_{1}}
-\frac{n+1}{t_{1}a_{n}}\frac{\partial a_{n}}{\partial t_{-1}}
-2a_{n}(a_{n}-1)
+\frac{(n+1)^{2}}{2t_{1}t_{-1}}
\frac{a_{n}-1}{a_{n}}
\nonumber \\
& &
+
\frac{1}{a_{n}}\frac{\partial a_{n}}{\partial t_{1}}
\frac{\partial a_{n}}{\partial t_{-1}}
+
\frac{1}{2}\frac{t_{1}}{t_{-1}}
\frac{1}{(a_{n}-1)a_{n}}
(\frac{\partial a_{n}}{\partial t_{1}})^{2}.
\label{k1}
\end{eqnarray}
In the same way, we eliminate $b_{n}^{*}$
 using (\ref{ww2}) and (\ref{w1})
and obtain an ODE for $a_{n}$
\begin{eqnarray}
\frac{\partial ^{2}a_{n}}{\partial t_{1}\partial t_{-1}}
&=&
\frac{n+1}{t_{1}a_{n}}\frac{\partial a_{n}}{\partial t_{-1}}
-\frac{n+1}{t_{-1}a_{n}}\frac{\partial a_{n}}{\partial t_{1}}
-2a_{n}(a_{n}-1)
+\frac{(n+1)^{2}}{2t_{1}t_{-1}}
\frac{a_{n}-1}{a_{n}}
\nonumber \\
& &
+
\frac{1}{a_{n}}\frac{\partial a_{n}}{\partial t_{1}}
\frac{\partial a_{n}}{\partial t_{-1}}
+
\frac{1}{2}\frac{t_{-1}}{t_{1}}
\frac{1}{(a_{n}-1)a_{n}}
(\frac{\partial a_{n}}{\partial t_{-1}})^{2}.
\label{k2}
\end{eqnarray}
The equality of (\ref{k1}) and (\ref{k2}) 
implies that
\begin{equation}
t_{1}\frac{\partial a_{n}}{\partial t_{1}}
=
t_{-1}\frac{\partial a_{n}}{\partial t_{-1}}
\end{equation}
Also this constraint  can be shown from 
(\ref{2}), (\ref{t1}) and (\ref{t2})  directly.
So $a_{n}$ are  functions of the radial coordinate
\begin{equation}
x=t_{1}t_{-1},
\end{equation}
only.
Then from (\ref{k1}) and (\ref{k2}) 
we can obtain
\begin{equation}
\frac{\partial ^{2}a_{n}}{\partial x^{2}}
=
\frac{1}{2}(\frac{1}{a_{n}-1}+\frac{1}{a_{n}})
(\frac{\partial a_{n}}{\partial x})^{2}
-\frac{1}{x}\frac{\partial a_{n}}{\partial x}
-\frac{2}{x}a_{n}(a_{n}-1)
+\frac{(n+1)^{2}}{2x^{2}}
\frac{a_{n}-1}{a_{n}}.
\label{p3}
\end{equation}
This  is an expression of the  Painlev\'{e}  V
equation (PV) with 
\begin{equation}
\alpha_{V}=0,\;\;
\beta_{V}=-\frac{(n+1)^{2}}{2},\;\;
\gamma_{V}= 2,\;\;
\delta_{V}=0.
\end{equation}
(\ref{p3}) is related to the usual one 
through 
\begin{equation}
a_{n} \longrightarrow c_{n}=\frac{a_{n}}{a_{n}-1}.
\end{equation}
(\ref{p3})  is  the  Painlev\'{e} III
 equation (P III) with 
(see \cite{o})
\begin{equation}
\alpha_{III}=\\  4(n+1)\;\;
\beta_{III}=-4n,\;\;
\gamma_{III}=4,\;\;
\delta_{III}=-4.
\end{equation}


\setzero
\section{Symmetric and Anti-symmetric model}

In this section we consider  reduced unitary matrix models.
The following reductions of the time variables $t_{k}$ leads to 
the symmetric and the anti-symmetric model:
\begin{equation} 
t_{k}=t_{-k}=t^{+}_{k}\;\;\;k=1,2,\cdots,\;\;\;({\rm symmetric\;\;model}) 
\end{equation}
and 
\begin{equation}
t_{k}=-t_{-k}=t^{-}_{k}\;\;\;k=1,2,\cdots,
\;\;\;({\rm anti-symmetric \;\;model}) 
\end{equation}

If $t_{1}^{+}$ are free variables while $t_{2}^{+}=t_{3}^{+}
=\cdots=0$,  
from (\ref{2}) $S_{n}=S_{n}^{*}$.
From (\ref{1}) and (\ref{2}) the string equation becomes
\begin{equation}
(n+1)S_{n}=t_{1}^{+}(S_{n+1}+S_{n-1})(1-S_{n}^{2}).
\label{sse}
\end{equation}
This is called the discrete Painlev\'{e} II (dP II) equation.
Appropriate continuous limit of (\ref{sse})
yields  the Painlev\'{e} II (P II) equation.
From (\ref{11}) and (\ref{22}) we can obtain the modified Volterra 
equation:\cite{m2}
\begin{equation}
\frac{\partial S_{n}}{\partial t_{1}^{+}}
=
-(1-S_{n}^{2})(S_{n+1}-S_{n-1}).
\label{mve}
\end{equation}
Appropriate continuous limit of (\ref{mve})
yields  the modified KdV equation.

(\ref{sse}) and (\ref{mve}) can be written in the form
\begin{equation}
2S_{n+1}=\frac{1}{1-S_{n}^{2}}
(\frac{n+1}{t_{1}^{+}}S_{n}-
\frac{\partial S_{n}}{\partial t_{1}^{+}}),
\label{a}
\end{equation}
and 
\begin{equation}
2S_{n-1}=\frac{1}{1-S_{n}^{2}}
(\frac{n+1}{t_{1}^{+}}S_{n}+
\frac{\partial S_{n}}{\partial t_{1}^{+}}),
\label{b}
\end{equation}
Writing (\ref{b}) as
\begin{equation}
2S_{n}=\frac{1}{1-S_{n+1}^{2}}
(\frac{n+2}{t_{1}^{+}}S_{n+1}+
\frac{\partial S_{n+1}}{\partial t_{1}^{+}}),
\label{c}
\end{equation}
and using (\ref{a}) to eliminate $S_{n+1}$ we obtain 
a second order ODE for $S_{n}$.
\begin{equation}
\frac{\partial^{2} S_{n}}{\partial t_{1}^{+2}}
=
-\frac{S_{n}}{1-S_{n}^{2}}
(\frac{\partial S_{n}}{\partial t_{1}^{+}})^{2}
-\frac{1}{t_{1}^{+}}\frac{\partial S_{n}}{\partial t_{1}^{+}}
+\frac{(n+1)^{2}}{t_{1}^{+2}}
\frac{S_{n}}{1-S_{n}^{2}}
-4S_{n}(1-S_{n}^{2}).
\end{equation}
It is important  to keep in mind
that  the relevant  function is $1-S_{n}^{2}=a_{n}$.
$a_{n}$ satisfies 
\begin{equation}
\frac{\partial ^{2}a_{n}}{\partial t_{1}^{+2}}
=
\frac{1}{2}(\frac{1}{a_{n}-1}+\frac{1}{a_{n}})
(\frac{\partial a_{n}}{\partial t_{1}^{+}})^{2}
-\frac{1}{t_{1}^{+}}\frac{\partial a_{n}}{\partial t_{1}^{+}}
-8a_{n}(a_{n}-1)
+\frac{2(n+1)^{2}}{t_{1}^{+2}}
\frac{a_{n}-1}{a_{n}}.
\label{p32}
\end{equation}
If we set $x=(t_{1}^{+})^{2}$, (\ref{p32}) 
is the same as  (\ref{p3}).
Then, we obtain (\ref{p32}) which is the special case of 
Painlev\'{e} III.
In conclusion, coupling the modified Volterra and 
the dP II, we can obtain the P III.
The double limit
\begin{equation}
n\rightarrow\infty,\;\;\;t_{1}^{+}\rightarrow\infty,\;\;\;
\frac{t^{+2}_{1}}{n}=O(1),
\end{equation}
maps
P III(\ref{p32}) to P II.
Clearly, this kind of limit can be 
discussed independently of the connection  with 
the modified Volterra and the modified KdV equation.

Next we consider the anti-symmetric model.
If $t_{1}^{-}$ are free variables while $t_{2}^{-}=t_{3}^{-}
=\cdots=0$,  
 from (\ref{2})
\begin{equation}  
S_{n}S_{n-1}^{*}=S_{n}^{*}S_{n-1}.
\label{2.5}
\end{equation}
From (\ref{1}) and (\ref{2})  the string equations
 become
\begin{eqnarray}
(n+1)S_{n}&=&t_{1}^{-}(-S_{n+1}+S_{n-1})(1-S_{n}S_{n}^{*}),
\nonumber \\
(n+1)S_{n}^{*}&=&t_{1}^{-}(S_{n+1}^{*}-S_{n-1}^{*})(1-S_{n}S_{n}^{*}).
\label{as1}
\end{eqnarray}
On the other hand, from (\ref{11}) and (\ref{22})
 we obtain the discrete nonlinear Schr\"{o}dinger (NLS) equation:\cite{m3}
\begin{eqnarray}
\frac{\partial S_{n}}{\partial t_{1}^{-}}
&=&
-(1-S_{n}S_{n}^{*})(S_{n+1}+S_{n-1}),
\nonumber \\
\frac{\partial S_{n}^{*}}{\partial t_{1}^{-}}
&=&
(1-S_{n}S_{n}^{*})(S_{n+1}^{*}+S_{n-1}^{*}).
\label{as2}
\end{eqnarray}
Using the same method in the symmetric model case 
we can obtain the P III.
Coupling the discrete NLS and the string equation,
 we can obtain P III.

Through the  transformation 
\begin{equation}
z\rightarrow iz\;\;\;it_{1}^{-}\rightarrow t_{1}^{+},
\label{tf}
\end{equation}
the anti-symmetric model is transformed into     the  symmetric model.
Then we get the B\"{a}cklund transformation  
from the discrete NLS to the modified Volterra equation:
\begin{equation}
S_{n}\longrightarrow (i)^{n+1}S_{n}.
\end{equation}
However we restrict $t_{1}^{-}$ to a real number,
 we can not transform the anti-symmetric model 
into the symmetric model.

We  change variables $a_{n}\rightarrow u_{n}=\ln a_{n}$.
Then (\ref{t1}) and (\ref{t2}) become 
\begin{equation}
\frac{\partial^{2} u_{n}}{\partial t_{1}\partial t_{-1}}
=
e^{u_{n+1}}-2e^{u_{n}}+e^{u_{n-1}}.
\label{te}
\end{equation}
In the anti-symmetric model 
from (\ref{2.4}) and (\ref{2.5})
we can get
\begin{eqnarray}
S_{n}&=&S_{n}^{*},\;\;\;\;a_{n}=1-S_{n}^{2},\;\;\;\;\;\;\;(n={\rm odd}),
\nonumber \\
S_{n}&=&-S_{n}^{*},\;\;\;\;a_{n}=1+S_{n}^{2},\;\;\;\;\;\;\;(n={\rm even}).
\label{sid}
\end{eqnarray}
In the case that $t_{1}^{-}$ is real, 
  we can see the oscillation of $a_{n}$.
This phenomenon can be seen only in the anti-symmetric model.

Here we consider the continuum limit near the anti-symmetric model.
We are interested in 
 $S_{n}^{2}=\epsilon g_{n}$,  $\epsilon\rightarrow 0$ 
 and $n\rightarrow \infty$.
We assume $
t_{1}=-t_{-1}+2\epsilon/n$ and define $g_{n+1}-g_{n}=\epsilon g_{n}'$ .

Then the continuum limit yields
\begin{equation}
u\equiv u_{n}=-u_{n+1}+\epsilon
(\pm g'_{n}+g^{2}_{n})+O(\epsilon^{2}),
\end{equation}
where 
$\pm$  corresponds to $n=$odd and $n=$even respectively.
So in the continuous limit 
(\ref{te}) becomes well known the 1D sinh Gordon equation
\begin{equation}
\frac{\partial^{2} u}{\partial t_{1}\partial t_{-1}}
=
-2\sinh 2u,
\end{equation}
where $t_{1}=-t_{-1}$.
Here we introduce the radial coordinate 
\begin{equation}
r=\sqrt{-t_{1}t_{-1}}.
\end{equation}
$u$ obeys an ODE of the form 
\begin{equation}
\frac{{\rm d}^{2} u}{{\rm d} r^{2}}
+\frac{1}{r}\frac{{\rm d} u}{{\rm d}r}
=
2\sinh 2u.
\end{equation}
This is the   P III with
\begin{equation}
\alpha_{III}=0\;\;
\beta_{III}=0,\;\;
\gamma_{III}=1,\;\;
\delta_{III}=-1.
\end{equation}
This equation is  obtained from the 2 states Toda  field equation, too.\cite{n}
Because of the oscillation of  $a_{n}$, in the continuous limit
 $u_{n}$ looks like having 2 states. 

At last we consider the relation between 
the symmetric and anti-symmetric  model 
from the determinant form.
The partition function of the symmetric model is 
\begin{equation}
\tau_{N}^{+}={\rm det}_{ij}I_{i-j}(t_{1}^{+}),
\end{equation}
where $I_{m}$ is the modified Bessel function of 
order $m$.\cite{ed}
In the same way, we can calculate the partition function of
the anti-symmetric model:
\begin{equation}
\tau_{N}^{-}={\rm det}_{ij}J_{i-j}(t_{1}^{-}),
\end{equation}
where  $J_{m}$ is the  Bessel function of 
order $m$.
(\ref{tf}) is also the transformation between 
the Bessel and the modified Bessel function.
(\ref{sid}) comes from the oscillation of the Bessel function.

\setzero
\section{Concluding remarks}
We try to reformulate 
the full unitary matrix model 
from the view points of 
integrable equations and string equations.
Coupling the Toda equation and the string equations,
we obtain the P III equation.
Because of  the Virasoro constraint,   
 $t_{1}$ and $t_{-1}$ have the symmetry.
This symmetry is like  complex conjugate.
Then we can use the radial coordinate.
This PIII also describe the phase transition 
between the week and strong coupling region.
Next  we consider the relation among the symmetric, anti-symmetric 
model and the P III equation.
If $t_{1}^{-}$ is a purely imaginary number, the anti-symmetric model can be transformed into  the anti-symmetric model.
Using this map we construct the B\"{a}cklund  
transformation from the discrete nonlinear Schr\"{o}dinger equation
to the modified Volterra equation  .
This map is also the transformation between the 
Bessel and the modified Bessel function.
If we restrict $t_{1}^{-}$ to a real number, 
 the symmetric and the anti-symmetric are  different.

\end{document}